\documentclass[twocolumn]{aastex61}

\newcommand\gsim{\,\lower3pt\hbox{$\sim$}\llap{\raise2pt\hbox{$>$}}\,}
\newcommand\lsim{\,\lower3pt\hbox{$\sim$}\llap{\raise2pt\hbox{$<$}}\,}

\usepackage{graphicx}
\usepackage{epsfig}
\usepackage{subfigure}
\usepackage{natbib}

\received{July 18, 2017}
\revised{September 11, 2017}
\accepted{September 21, 2017}
\submitjournal{Astrophysical Journal}

\shortauthors{LUGAZ ET AL.}

\shorttitle{CME EXPANSION AND SHOCKS}

\begin{document}

%% ------------------------------------------------------------------------ %%
%
%  TITLE
%
%% ------------------------------------------------------------------------ %%

\title{Importance of CME Radial Expansion on the Ability of Slow CMEs to Drive Shocks}

\author[0000-0002-1890-6156]{No{\'e}\ Lugaz}
\affiliation{Space Science Center, Institute for the Study of Earth, Oceans, and Space, University of New Hampshire, Durham, NH, USA}
\affiliation{Department of Physics, University of New Hampshire, Durham, NH, USA}
\author{Charles~J. Farrugia}
\affiliation{Space Science Center, Institute for the Study of Earth, Oceans, and Space, University of New Hampshire, Durham, NH, USA}
\affiliation{Department of Physics, University of New Hampshire, Durham, NH, USA}
\author{Reka~M. Winslow}
\affiliation{Space Science Center, Institute for the Study of Earth, Oceans, and Space, University of New Hampshire, Durham, NH, USA}
\author{Colin~R.~Small}
\affiliation{Department of Physics, University of New Hampshire, Durham, NH, USA}
\author{Thomas Manion}
\affiliation{Department of Physics, University of New Hampshire, Durham, NH, USA}
\author{Neel~P. Savani}
\affiliation{NASA/GSFC and University of Maryland Baltimore County, Greenbelt, MD, USA}

%% ------------------------------------------------------------------------ %%
%
%  ABSTRACT
%
%% ------------------------------------------------------------------------ %%

\begin{abstract}
Coronal mass ejections (CMEs) may disturb the solar wind either by overtaking it, or by expanding into it, or both. CMEs whose front moves faster in the solar wind frame than the fast magnetosonic speed, drive shocks. Such shocks are important contributors to space weather, by triggering substorms, compressing the magnetosphere and accelerating particles. In general, near 1~AU, CMEs with speed greater than about 500~km\,s$^{-1}$ drive shocks, whereas slower CMEs do not. However, CMEs as slow as 350~km\,s$^{-1}$ may sometimes, although rarely, drive shocks. Here, we study these slow CMEs with shocks and investigate the importance of CME expansion in contributing to their ability to drive shocks and in enhancing shock strength. Our focus is on CMEs with average speeds under 375~km\,s$^{-1}$. From Wind measurements from 1996 to 2016, we find 22 cases of such shock-driving slow CMEs, and, for about half of them (11 out of the 22), the existence of the shock appears to be strongly related to CME expansion. We also investigate the proportion of all CMEs with speeds under 500~km\,s$^{-1}$ with and without shocks in solar cycles 23 and 24, depending on their speed. We find no systematic difference, as might have been expected on the basis of the lower solar wind and Alfv{\'e}n speeds reported for solar cycle 24 vs.\ 23. The slower expansion speed of CMEs in solar cycle 24 might be an explanation for this lack of increased frequency of shocks, but further studies are required.

\end{abstract}

\section{INTRODUCTION} \label{intro}

Fast forward shocks measured near 1~AU are primarily driven by coronal mass ejections (CMEs) and corotating interaction regions (CIRs). Aside from their drivers, shocks, by themselves, can elicit significant geo-effects, as is clear from their association with sudden storm commencements and sudden impulses \citep[]{Nishida:1964, Siscoe:1968},  and triggered substorms \citep[]{Schieldge:1970,Kokubun:1977}. Specifically, sudden impulses are one of the leading causes of large geomagnetically induced currents \citep[]{Kappenman:2003,Carter:2015}. In addition, shocks are one of the primary accelerators of solar energetic particles \citep[e.g., see][]{Reames:1999} and the sheath regions downstream of shocks, which contains magnetic fields compressed by the shock, can drive intense geomagnetic storms \citep[]{Tsurutani:1988,Lugaz:2016b}. While all of these have been well studied, there have been relatively fewer investigations looking at the reasons why some CMEs drive shocks, whereas others do not. This may appear as a relatively straightforward question, but it involves a combination of CME properties (speed, expansion) and solar wind characteristics (speed and fast magnetosonic speed). 

Statistical studies of shock properties over several years include \citet{Berdichevsky:2000}, \citet{Oh:2007}, \citet{Vorotnikov:2008} and \citet{Kilpua:2015}. Such studies have shown that about 25\% of CIRs and 50\% of CMEs drive a shock at 1 AU \citep[]{Jian:2006,Jian:2006b}. For a shock to form, the CME or CIR must be faster, in the solar wind frame, than the fast magnetosonic speed. Because the magnetosonic speed depends on the shock angle, $\theta$, its upper value, for $\theta = 90^\circ$ is often used. CMEs can drive shocks due to their fast propagation speed as compared to the background solar wind speed, or because they expand into it. The first behavior has been well studied, whereas the second one, usually referred to as ``over-expansion'' has only be reported at high-latitudes \citep[]{Gosling:1994}. Note that \citet{Siscoe:2008} wrote in detail about the difference between a propagation and an expansion sheaths in front of magnetic ejecta. That study was primarily concerned with the non-radial expansion of CMEs, associated with the fact that CMEs keep a more or less constant cone angle throughout their heliospheric propagation \citep[]{Riley:2004b}. In the present study, we distinguish between the formation of a shock and sheath region associated with the CME radial expansion to that associated with the CME bulk propagation motion.

The focus of this paper is on shocks being driven by relatively slow CMEs and on the contribution of CME radial expansion to their shock driving ability. \citet{Poedts:2016} recently reported a numerical simulation of a slow CME ($\sim 400$~km\,s$^{-1}$) driving a shock in the interplanetary space. The authors found that the simulated slow CME with a speed at 1~AU of 390~km\,s$^{-1}$ drives a shock starting at 0.25~AU. The CME expansion speed was found to be about 40~km\,s$^{-1}$ and the CME central speed was only 20~km\,s$^{-1}$ faster than the solar wind. Thus, without expansion, this CME would not have been able to drive a shock. \citet{Liu:2016} reported on three slow CMEs in solar cycle 24 with speed under 400~km\,s$^{-1}$, two of which were nonetheless preceded by a shock at 1~AU. \citet{Gopalswamy:2010} reported on interplanetary shocks measured near 1~AU which were associated with radio-quiet CMEs ({\it i. e.}\ CMEs that do not drive shocks that accelerate electrons in the corona), finding that about one-third of shocks were associated with radio-quiet CMEs. This indicates that these shocks formed high in the corona or beyond. These shocks were typically slower and weaker and associated with slower CMEs. The authors considered that the variation in Alfv{\'e}n speed in the inner heliosphere may explain why some slow CMEs are radio-loud whereas some faster CMEs are radio-quiet.

\begin{table*}[ht]
\centering
\begin{tabular}{|c|c|c|c|c|c|c|c|c|}
\hline
Event & Date & $t_\mathrm{shock}$ & Sheath $\Delta t$ (h) & CME Start & CME End & $B_\mathrm{CME\, max}$ (nT) & $V_\mathrm{CME}$ (km\,s$^{-1}$) & $\theta_\mathrm{Bn}$\\
\hline
\multicolumn{9}{|c|}{{\bf Average Mach Less than 1 and Front Mach Greater than 1}}\\
\hline
1 & 1998/08/19 & 18:40 & 11 & 08/20 06:00 & 08/21 20:00 & 16.5 & 320 & 51\\
2 & 2001/10/31 & 13:47 & 8.2 & 10/31 20:00 & 11/02 12:00 & 13 & 340 & 14\\
3 & 2009/02/03 & 19:21 & 4.6 & 02/04 00:00 & 02/04 16:00 & 11.5 & 360 & 74\\
4 & 2010/05/28 & 01:53 & 17.1 & 05/28 19:00 & 05/29 17:00 & 14.5 & 355 & 28\\
5 & 2012/09/30 & 22:19 & 1.7 & 10/01 00:00 & 10/02 00:00 & 21.5 & 370 & 67\\
\hline
\multicolumn{9}{|c|}{{\bf Average Mach and Front Mach Less than 1}}\\
\hline
6 & 1997/12/30 & 01:14 & 8.8 & 12/30 10:00 & 12/31 11:00 & 14 & 365 & 60\\
7 & 1998/03/04 & 11:03 & 1.8 & 03/04 13:00 & 03/06 09:00 & 12.5 & 345 & 35\\
8 & 2000/07/26 & 19:00 & 7 & 07/27 02:00 & 07/28 02:00 & 8 & 355 & 88\\
9 & 2001/04/21 & 15:29 & 8.8 & 04/21 23:00 & 04/23 03:00 & 15 & 365 & 52\\
10 & 2011/03/29  & 15:10 & 9.1 & 03/29 23:00 & 03/31 04:00 & 14.5 & 365 & 77\\
11 & 2014/02/15 & 12:47 & 15.4 & 02/16 05:00 & 02/16 16:00 & 17.5 & 375 & 88\\
\hline
\multicolumn{9}{|c|}{{\bf Average Mach and Front Mach Greater than 1}}\\
\hline
12 & 1997/05/26 & 09:09 & 7 & 05/26 16:00 & 05/27 10:00 & 11 & 335 & 57 \\
13 & 1997/12/10 & 04:33 & 14.1 & 12/10 18:40 & 12/12 00:00 & 16 & 350 & 69 \\
14 & 2002/03/18 & 13:14 & 17.2 & 03/19 05:00 & 03/20 16:00 & 21.5 & 370 & 39\\
15 & 2012/10/31 & 14:28 & 9.1 & 11/01 00:00 & 11/02 03:00 & 15.5 & 350 & 86\\
16 & 2014/08/19 & 05:49 & 10.2 & 08/19 16:00 & 08/21 05:00 & 21 & 360 & 85 \\
17 & 2016/01/18 & 21:21 & 12.7 & 01/19 10:00 & 01/21 00:00 & 17 & 360 & 55\\
\hline
\multicolumn{9}{|c|}{{\bf No Clear Expansion}}\\
\hline
18 & 1998/06/13 & 19:18 & 8.6 & 06/14 04:00 & 06/15 06:00 & 12 & 345 & 32\\
19 & 2010/02/10 & 23:58 & 8.5 & 02/11 08:00 & 02/12 03:00 & 8.5 & 360 & 61\\
20 & 2012/01/21 & 04:02 & 2 & 01/21 06:00 & 01/22 08:00 & 12.5 & 330 & 81\\
21 & 2012/09/30 & 10:15 & 3.8 & 09/30 14:00 & 09/30 20:00 & 8 & 315 & 77\\
22 & 2016/11/09 & 05:45 & 18.2 & 11/10 00:00 & 11/10 16:00 & 13 & 360 & 88 \\
\hline
\end{tabular}
\caption{List of 22 CMEs with average speed less or equal to 375~km\,s$^{-1}$ that drove a shock. All times are in UT. The $B_\mathrm{CME\, max}$ column lists the maximum magnetic field strength in the magnetic ejecta,  $V_\mathrm{CME}$ lists the average magnetic ejecta solar wind speed and $\theta_\mathrm{Bn}$ is the angle between the shock normal and the upstream magnetic field.}
\label{twentytwo}
\end{table*}

Regarding CME expansion and its relationship with shocks, the main focus of research on this topic has been on shock pairs driven by ``over-expanding'' CMEs at high latitudes as measured by Ulysses \citep[]{Gosling:1994,Gosling:1998,Reisenfeld:2003}. The authors of these studies concluded that a high internal pressure inside the magnetic ejecta as compared to the ambient solar wind and a speed comparable to that of the solar wind are the two necessary conditions for a CME to form an ``over-expanding'' shock. Under this explanation, there is no true reason why these shock pairs could not form close to the ecliptic.  In fact, \citet{Gosling:1995} noted that ``[i]t is not  known  why  shock pairs driven by  over-expansion are restricted to  high latitudes.'' Note that \citet{Manchester:2006} proposed an alternative origin for these shocks. In their view, the pair of shocks forms due to the interaction of the CME with the bimodal solar wind: the CME latitudinal expansion results in the interaction between the deflected slow solar wind in front of the CME and the deflected fast wind behind the CME, creating a reverse shock.  \citet{Riley:2006} gave some counter-arguments in favor of the original interpretation by \citet{Gosling:1994}. In general, CME radial expansion is thought to occur sub-Aflv{\'e}nically; this result dates back from early studies in the 1980s that found that CME expansion is in general close to half the Alfv{\'e}n speed of the medium surrounding the CME \citep[]{Klein:1982,Burlaga:1982}. 

Lastly, it has already been well studied that solar cycle 24 (2008--present) is weak in many respect, including solar wind characteristics, number of CMEs measured near Earth and geo-effects \citep[]{Gopalswamy:2015}. Of particular relevance to this study, the solar wind speed and interplanetary magnetic field have been low when compared to the previous solar cycle. Our interest is to determine whether these changes are reflected in the association of shocks with slow CMEs near 1 AU. 

The rest of the article is organized as follows: in section \ref{overall}, we give some general results about the 22 slowest shock-driving CMEs in solar cycles 23 and 24. In section~\ref{example}, we present four examples of slow CMEs, three of which are driving a shock. In section \ref{data}, we present general considerations regarding shocks driven by CMEs for solar cycle 23 and 24 and compare these expectations to the actual number of CME-driven shocks. We summarize and conclude in section \ref{conclusion}.

\begin{table*}[ht]
\centering
\begin{tabular}{|c|c|c|c|c|c|c|c|c|c|c|c|c|c|}
\hline
Event & $V_\mathrm{exp}$ & $V_\mathrm{sw}$ & $V_\mathrm{ms}$ & $V_\mathrm{cme}$ & $V_\mathrm{front}$ & $V_\mathrm{max}$ & $V_\mathrm{shock}$  & $M_\mathrm{cme}$ & $M_\mathrm{front}$  & $M_\mathrm{max}$ & $M_\mathrm{shock}$ & $V_\mathrm{exp}/V_\mathrm{cme}$ &  $\xi$ \\
\hline
1 & 32 & 275 & 45 & 320 & 350 & 350 & 325 & 0.99 & 1.65 & 1.65 & 1.8 & 0.10 & 0.92\\
2 & 42 & 320 & 57 & 345 & 400 & 400 & 415 & 0.35 & 1.2 & 1.2 & 2.1 & 0.12 & 1.26\\
3 & 27 & 305 & 65 & 360 & 400 & 400 & 375 & 0.83 & 1.4 & 1.4 &  1.2 & 0.075 & 1.85\\
4 & 25 & 315 & 48 & 355 & 390 & 390 & 295 & 0.84 & 1.6 & 1.6 & 1.4 & 0.070 & 0.53\\
5 & 35 & 320 &  58 & 370 & 400 & 420 & 450 & 0.86 & 1.4 & 1.7 &  2.9 & 0.095 & 0.79\\ 
\hline
6 & 0 & 320 & 58 & 365 & 375 & 410 & 370 & 0.78 & 0.95 & 1.6 & 1.95 & 0 & 1.1\\
7 & 40 & 350 & 51 & 345 & 390 & 390 & 440 & -0.10 & 0.78 & 0.78 & 2.2 & 0.12 & 0.72\\
8 & 20 & 350 & 50 & 360 & 370 & 400 & 415 & -0.10 & 0.20 & 0.81 & 1.3 & 0.056 & 0.82\\
9 & 30 & 360 & 46 & 365 & 390 & 415 & 370 & 0.11 & 0.65 & 1.2 & 1.6 & 0.082 & 0.89\\
10 & 25 & 330 & 63 & 365 & 385 & 405 & 395 & 0.56 & 0.87 & 1.2 & 1.8 & 0.068 & 0.6\\
11 & 32 & 375  & 70 & 375 & 405 & 450 & 470 & 0 & 0.43 & 1.1 & 2 & 0.085 & 1.08\\
\hline
12 & 20 & 285 & 41 & 335 & 360 & 360 & 335 & 1.2 & 1.8 & 1.8 & 1.4 & 0.060 & 0.80 \\
13 & 20 & 285 & 60 & 350 & 375 & 390 & 390 & 1.1 & 1.5 & 1.7 & 2.2 & 0.057 & 0.57\\
14 & 42 & 295 & 52 & 370 & 420 & 480 & 515 & 1.4 & 2.4 & 3.6 & 4.1 & 0.11 & 2.7\\
15 & 30 & 290 & 45 & 350 & 370 & 380 & 390 & 1.3 & 1.8 & 2.0 & 2.2 & 0.086 & 0.78\\
16 & 45 & 295 & 51 & 360 & 415 & 430 & 270 & 1.3 & 2.4 & 2.7 & 1.3 & 0.12 & 1.03\\
17 & 15 & 300 & 58 & 360 & 385 & 385 & 355 &  1.05 & 1.5 & 1.5 & 1.9 & 0.042 & 0.32\\
\hline
18 & 0 & 315 & 64 & 345 & 350 & 385 & 270 & 0.47 & 0.55 & 1.1 & 1.7 & 0 & - \\
19 & 2 & 300 & 53 & 360 & 365 & 380 & 315 & 1.1 & 1.2 & 1.5 & 1.8 & 0.0056 & - \\
20 & 2 & 305 & 56 & 330 & 330 & 330 & 335 & 0.45 & 0.45 & 0.45 & 1.4 & 0.0060 & -\\
21 & -5 & 270 & 38 & 315 & 310 & 320 & 350 & 1.2 & 1.05 & 1.3 & 2.0 & -0.016 & -0.6\\
22 & -15 & 295 & 50 & 360 &  345 & 375 & 360 & 1.3 & 1 & 1.6 & 1.4 & -0.042 & -0.58\\
\hline
\end{tabular}
\caption{Speeds, Mach numbers and expansion of the CMEs and shocks from Table~\ref{twentytwo}. The columns list from left to right, the event number, expansion speed, upstream solar wind speed and fast magnetosonic speed, the average speed of the magnetic ejecta, the speed at the front of the magnetic ejecta, the maximum speed in the CME, including the sheath, the shock speed in a rest frame, the Mach numbers associated with the previous four speeds, the ratio of the expansion speed to the average speed and the dimensionless $\xi$ expansion parameter of \citet{Demoulin:2008}. All speeds are in km\,s$^{-1}$ and all other numbers are dimensionless.}
\label{Mach}
\end{table*}

\section{Characteristics of the Slowest 22 Shock-Driving CMEs}\label{overall}

We identify the slowest CMEs that drove a shock from 1996 to 2016 by combining the ICME database of \citet{Richardson:2010} and the shock database of \citet{Kilpua:2015}. For each of these CMEs, we visually inspected the Wind 3-second magnetic field measurements from the MFI instrument and the $\sim\,100$-second plasma measurements from the SWE instrument. For each CME with a reported speed less or equal to 380~km\,s$^{-1}$ from the database of \citet{Richardson:2010}, we used these Wind data to calculate the average speed in the magnetic ejecta, the expansion speed, the magnetic ejecta front speed, the maximum speed (including the sheath), and the dimensionless $\xi$ expansion parameter following \citet{Demoulin:2008}. We determined the presence of a shock using the database of \citet{Kilpua:2015} and identifying shocks within 2 hours of the disturbance start time listed in \citet{Richardson:2010}. This disturbance start time is often a shock, or shock-like structure. We focus only on CMEs with average speed under 375~km\,s$^{-1}$. This speed threshold was chosen to focus on the slowest CMEs and to still obtain about 10 CME-driven shocks in solar cycles 23 and 24. 

For some CMEs, a difference of the average speed of 5--10~km\,s$^{-1}$ slower or faster than that reported by \citet{Richardson:2010} lead us to either include or exclude it from our list. We  excluded six additional events: two had sheaths longer than 30 hours, three had sheath durations longer than that of the magnetic ejecta and one corresponds to two shocks within 24 hours of the start of a slow magnetic ejecta. Such long-duration sheaths are unusual and the shock may be driven by a separate ejection that did not impact Earth. We are left with 22 cases of CMEs with average less or equal to  375~km\,s$^{-1}$ that drove a shock. The shock database lists the shock properties (shock angle, speed, normal direction, Mach number, fast upstream magnetosonic speed) and we use these directly from the database. Upstream quantities are averaged over 8 minutes (from 9 minutes to 1 minute before the shock). Shock normal estimates are known to come with significant uncertainties \citep[]{Szabo:1994} depending on the method chosen and the exact time windows chosen for the upstream and downstream conditions. By using the database of \citet{Kilpua:2015} where over 2,000 shocks are analyzed, we are left with a self-consistent set of values for the shocks in this study. In addition, we calculated the propagation, leading edge and maximum speed Mach numbers, denoted as $M_\mathrm{cme}$, $M_\mathrm{front}$ and $M_\mathrm{max}$, respectively. Each of these Mach numbers are calculated in the solar wind frame, using the fast magnetosonic speed upstream of the shock for $\theta = 90^\circ$ (maximum fast magnetosonic speed).  The propagation Mach number is the Mach number corresponding to the CME average speed in the solar wind frame. The leading edge Mach number includes the effects of expansion, when present. These three Mach numbers are calculated using the upstream solar wind speed using the formula, $M_i = (V_i - V_\mathrm{sw}) / V_\mathrm{ms}$, with $i$ representing the average, front or max speeds. The shock Mach number is derived using the solar wind speed in the shock's normal direction $M_\mathrm{shock} = (V_\mathrm{shock} - \vec{V_\mathrm{sw}}\cdot\vec{n}) / V_\mathrm{ms}$, where $\vec{n}$ is the shock normal. As such, when a shock is inclined with respect to the solar wind direction (typically radial), the shock Mach number can be significantly higher than the front Mach number, even for similar front and shock speeds. Using our definition, a shock with any inclination with respect to the solar wind direction and interplanetary magnetic field can exist for a front or maximum Mach number greater than 1. 

The two slowest events are the 1998 August 20 ejecta (average speed of 320~km\,s$^{-1}$ and maximum speed of 350~km\,s$^{-1}$) and the 2012 September 30 ejecta (average speed of 315~km\,s$^{-1}$, maximum speed of 320~km\,s$^{-1}$). The 2012 September 30 is part of a series of two shock-driving  and interacting slow CMEs that have been studied in details in \citet{Liu:2014b,Lugaz:2015b}. These shocks were essential in compressing Earth's magnetosphere and resulted in loss of energetic electrons from the outer radiation belts \citep[]{Turner:2014a}. 

We can overall think about these slow CMEs as belonging to one of four categories: (a) CMEs with $M_\mathrm{cme} < 1 < M_\mathrm{front}$. These CMEs do not have propagation speeds  fast enough to drive a shock but, with the additional speed from expansion, their leading edge is fast enough to drive a shock. There are five cases like this. This includes one for which the center Mach number is 0.99. (b) CMEs with $M_\mathrm{cme} < M_\mathrm{front} < 1 $. These CMEs have a propagation speed not fast enough to drive a shock, the expansion results in a CME front with a significantly faster speed than the center but still not enough to drive a shock and the sheath is typically slightly faster. Six cases are like this, four of which have a maximum speed fast enough to explain the existence of a shock. The other two cases can be understood when the shock angle is taken into consideration (such a CME could not drive a quasi-perpendicular shock). (c) CMEs with $1 < M_\mathrm{cme} < M_\mathrm{front}$. Six events fit this category. (d) CMEs with complex velocity structures, no clear expansion, sometimes being overtaken. Five events fit this category. 

Table~\ref{twentytwo} shows these 22 events, separated into these four categories. Table~\ref{Mach} lists the speeds and Mach numbers associated with these events. We consider that the events for which the average CME speed is not enough to drive a shock but expansion is significant (a and b, above, 11 events) are events for which expansion was central in driving a shock. For events belonging to category (c), it is impossible to quantify clearly how fundamental expansion is for driving such shocks: the propagation Mach number is of the order of 1.2 (maximum of 1.44) for these six events. Such weak Mach number may not be enough to drive a clear shock. On the other hand, when expansion is taken into consideration, the CME front Mach number is 1.90, relatively typical for CME-driven shocks at 1~AU. For all six events, the Mach number of the CME front is at least 0.4 higher than the average CME Mach. 

 %%%%%%%%%%%%%%%%%%%%%%%
\begin{figure*}[htb]
\centering
{\includegraphics*[width=8.5cm]{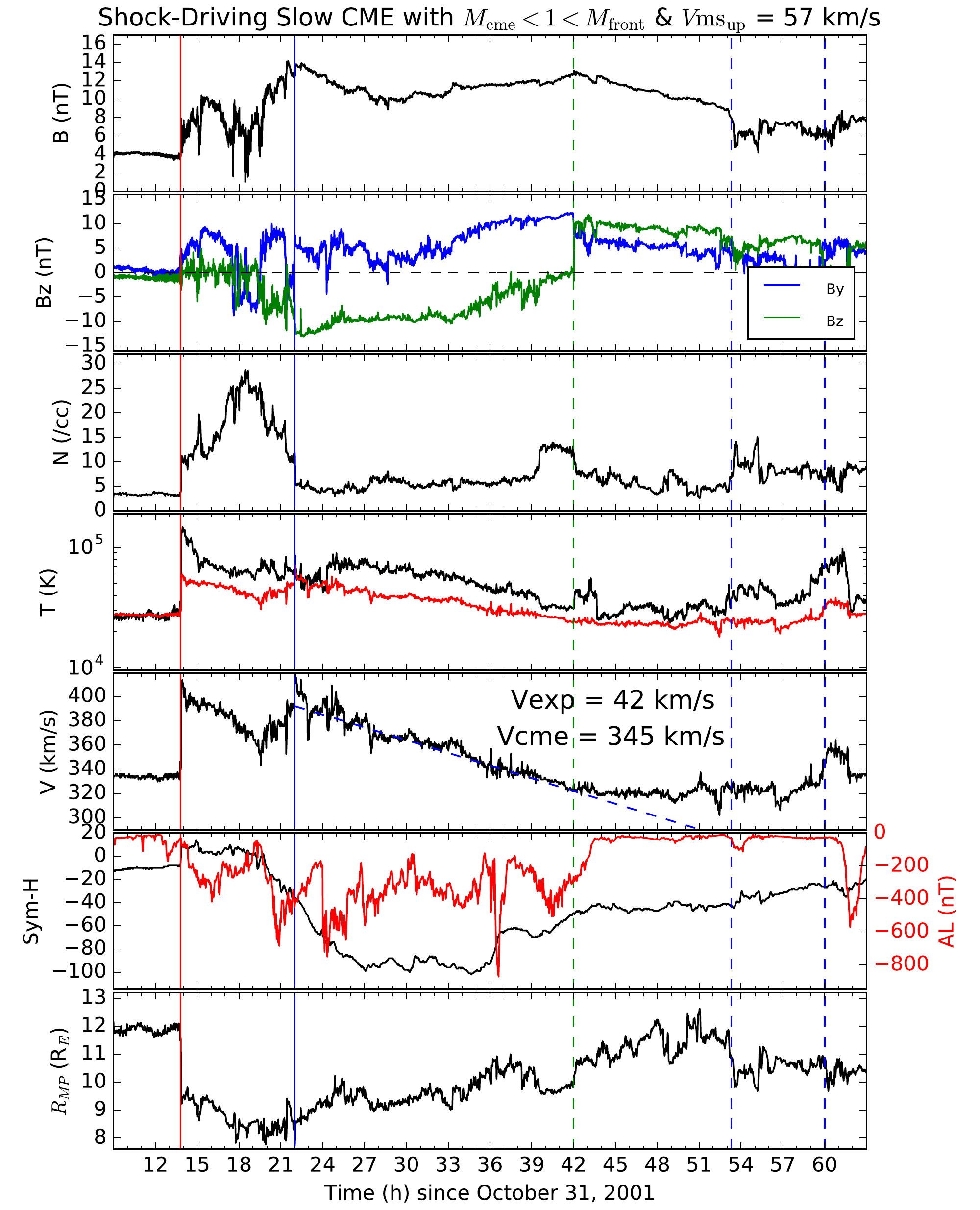}}
{\includegraphics*[width=8.5cm]{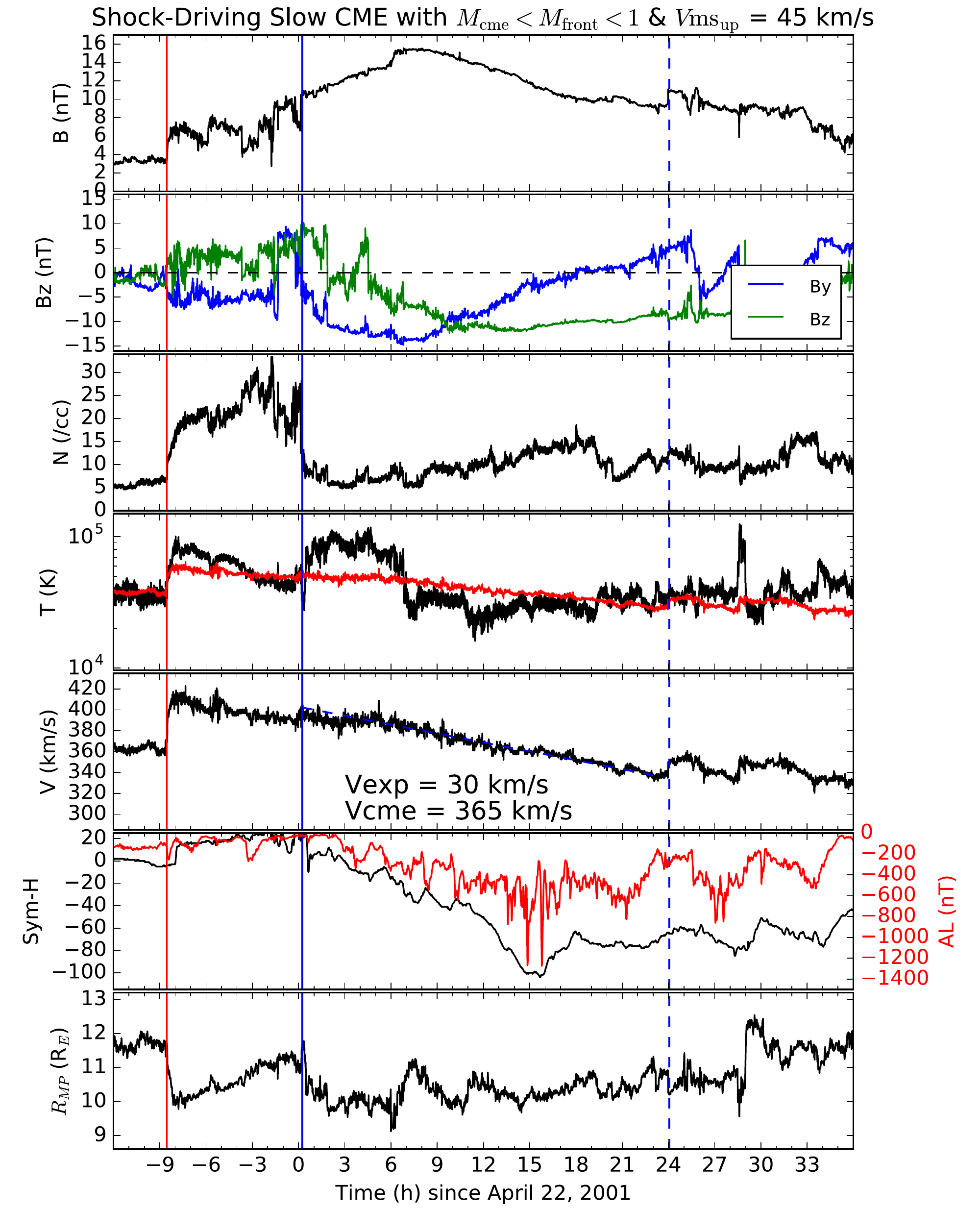}}
\caption{Two examples of a  shock-driving slow CME. For both CMEs, the panels show from top to bottom, the total magnetic field strength, the $y$ (blue) and $z$ (green) components of the magnetic field vector in GSE coordinates, the proton density, the temperature and expected temperature --red, following \citet{Lopez:1987}--, the velocity, the Sym-H (black) and AL (red) indices and the sub-solar magnetopause location following \citet{Shue:1998}. The left panels show the  2001 October 31 2001 CME; the right panels the 2001 April 22 CME. The shocks are marked with red vertical lines, the magnetic ejecta boundaries with blue or green lines with dash indicating one of several possibilities.}
\label{Oct2001}
\end{figure*}
%%%%%%%%%%%%%%%%%%%%%%%%

These 22 events occur in very slow solar wind (average of 312~km\,s$^{-1}$, 10 events occur with solar speed of 300~km\,s$^{-1}$ or less) and with low fast magnetosonic speed (average of 54~km\,s$^{-1}$). There is a range of expansion speeds and expansion parameters. Expansion parameters are mostly around the typical value of 0.8 but six CMEs are over-expanding in the sense that the dimensionless parameter is greater than 1. In all cases, the expansion itself is sub-fast with the typical expansion speed being about half of the upstream fast magnetosonic speed, confirming past studies \citep[]{Klein:1982}. The 2001 November 1 and 2014 November 20 CMEs are the ones with an expansion speed closest to the local fast magnetosonic speed. Five events have a shock speed lower than the CME speed; this occurs because the shock is oblique or quasi-parallel in all but one case. For the one exception, on 2014 August 19 CME, there is a strong positive speed gradient from the region just downstream of the shock to the front of the magnetic ejecta (by over 100~km\,s$^{-1}$) and a significant expansion in the magnetic ejecta by about 90~km\,s$^{-1}$. It is unclear to us what might cause such a behavior. 

\section{Examples of Shock-Driving Slow CMEs}\label{example}
We focus now on four specific cases of slow CMEs, three of them driving a fast forward shock and corresponding to the first three categories discussed above and one for a non-shock driving CME. 

\subsection{Example with $M_\mathrm{cme} < 1 < M_\mathrm{front}$: 2001 October 31 (Event \#2)}

On 2001 October 31 at 13:45~UT, a fast-forward shock impacted Wind. It had a speed of 415~km\,s$^{-1}$ and a fast magnetosonic Mach (hereafter, Mach) number of about 2.1. Eight hours later around 22~UT, a decrease in density and the beginning of a long-duration period of southward magnetic field marked the start of a magnetic ejecta. The left panel of Figure~\ref{Oct2001} shows the magnetic field, plasma and geomagnetic measurements associated with this CME. The maximum speed of the ejecta, reached at the front, was  392~km\,s$^{-1}$, close to the shock speed. As is often the case, determining the back boundary of the CME is not straight-forward \citep[e.g., see][]{Zurbuchen:2006,AlHaddad:2013}. This is because the end of elevated and smooth magnetic field rarely coincide with the end of the plasma characteristics of magnetic ejecta (low density and temperature) and the changes in charge state \citep[]{Richardson:2010}. In addition, shocks, discontinuities and current sheets are often observed towards the back of magnetic ejecta \citep[]{Collier:2007,Foullon:2007,Owens:2009,Lugaz:2015b}. \citet{Richardson:2010} list 12:00 UT on November 2 (last blue dashed line at 60 hours), the Wind ICME database lists 04:47 UT (first blue dashed line at 52.8 hours), the DREAMS database lists 15:15 on November 1 (39.25 hours). It is also possible to consider 18:00 UT on November 1 (green dashed line at 42 hours) as the end of the magnetic ejecta, as identified in this study and independently in \citet{Owens:2009}. The times around 39 and 42 hours correspond to the beginning and end of an increase in density. The end also corresponds to a sharp discontinuity in the magnetic field direction. The magnetic ejecta end at 52.8 hours correspond to the end of the smooth and elevated magnetic field, whereas the end time at 60 hours correspond to an increase in solar wind velocity. Hereafter, we use the end at 52.8 hours as the end of the magnetic ejecta but we also list values obtained by considering that the CME ends at 42 hours. Using the end time at 52.8 hours, the CME average speed is 344~km\,s$^{-1}$  and the center speed is 340~km\,s$^{-1}$.

The solar wind speed upstream of the shock was 320~km\,s$^{-1}$ and the fast magnetosonic speed was 57~km\,s$^{-1}$, so in most cases, this CME is not expected to drive a shock based on its central speed. This is because the fast Mach number of the CME center in the solar wind frame is 0.35. However, since the front speed is about 390~km\,s$^{-1}$, the CME front has a fast Mach number of 1.2, consistent with the presence of a shock. It can be clearly seen, then, that the expansion is essential in driving this shock wave. This shock wave compressed Earth's magnetosphere significantly (see bottom panel following \citet{Shue:1998}) from about 12 to 9~$R_E$. During the sheath passage, the main phase of an intense geomagnetic storm started. The Dst index reached $-60$~nT at the end of the sheath at 23:00~UT, indicating that the sheath itself was geo-effective. A small substorm (AL reached $-600$~nT) occurred during the sheath passage. 

Next, we consider how common this expansion was. The expansion speed is 42~km\,s$^{-1}$ using half the difference between the front and back CME velocities; this gives a $\Delta V_\mathrm{exp} / V_\mathrm{center}$ of 0.12. Using the slope of the velocity curve in the period when there is expansion $\frac{\Delta V}{\Delta t} = 0.97$~m\,s$^{-2}$, the dimensionless expansion parameter of \citet{Demoulin:2008} can be calculated as $\xi = \frac{D}{V_c^2} \frac{\Delta V}{\Delta t} = 1.26$. Here, $D$ is the distance where the measurement is made (here 1~AU) and $V_c$ is the CME center speed. Typical values of the dimensionless expansion parameters are 0.8 with values above 1 showing over-expansion \citep[]{Gulisano:2010}. 

Using an end time at 18:00 UT on November 1 (42 hours), the average and center speeds are 359~km\,s$^{-1}$ and the expansion speed is 33~km\,s$^{-1}$. The CME center is still sub-fast with a Mach  number of 0.67. The expansion gives $\Delta V_\mathrm{exp} / V_\mathrm{center} = 0.092$ or an expansion parameter of 1.13. This confirms that the CME is strongly expanding and that, without expansion, it would not drive a shock. 

We attempted to identify the associated eruption from LASCO measurements. There was  no full halo CME and only one partial halo from October 27 00 UT to October 30 00 UT ({\it i.\ e}.\ 38 to 110 hours before the shock arrival time). The partial halo from October 28 at 00:26 UT (85.5 hours before the shock arrival) was a slowly accelerating CME with a second-order leading edge speed around 24~$R_\odot$ of 750~km\,s$^{-1}$. Its appearance in LASCO images and associated EIT images indicate an eruption from the eastern limb, unlikely to have impacted Earth. As such, it is unlikely, but possible, that this is the eruption that resulted in the shock and magnetic ejecta at 1~AU. We deem it more likely though that, during this active period, a weak halo CME would not have been detected. 
 
 \subsection{Example with $M_\mathrm{cme} < M_\mathrm{front} < 1$: 2001 April 21 (Event \#9)}
The 2001 April 22 magnetic ejecta (Figure~\ref{Oct2001}, right panels) is similar in some respects to the 2001 October event, for example in that it is convected with the solar wind. The CME central speed of 365~km\,s$^{-1}$ is almost identical to the upstream solar wind of 360~km\,s$^{-1}$, meaning that the center Mach number was between 0 and 0.1. As such, as in the 2001 October CME discussed above, the CME without expansion was not fast enough to drive a shock. The CME expansion of about 30~km\,s$^{-1}$ results in a front speed of 395~km\,s$^{-1}$, which combined with the solar wind speed and the fast magnetosonic speed of 45~km\,s$^{-1}$ results in a leading edge Mach number of 0.8. It can therefore appear surprising that a shock precedes this magnetic ejecta and impacted Wind with a speed of 380~$\pm~20$~km\,s$^{-1}$ and a Mach number of 1.7 $\pm~0.1$. This is because the shock is inclined with respect to the upstream magnetic field direction (the angle between the shock normal and the magnetic field, $\theta_\mathrm{Bn} \sim 50^\circ$), which results in the solar wind speed normal to the shock's normal to be of the order of 300~km\,s$^{-1}$. 

The sheath in front of the magnetic ejecta has the same decreasing speed profile that the ejecta itself, with a maximum speed of 415~km\,s$^{-1}$ reached just behind the shock. Such a speed corresponds to a Mach of 1.2. This expanding sheath brings the question as to whether the shock is driven at 1~AU. No clear answer can be given since, on the one hand, the shock speed is comparable (and slightly lower) than the speed at the front of the magnetic ejecta, but on the other hand, the speed of the sheath is even higher. Our interpretation is that the CME expansion drove this shock in the inner heliosphere, but it may not be driven anymore as the CME passes Earth. Aging effects (the CME slows down as it propagates and the front hit Earth almost a day earlier than the back) could explain the speed profile rather than expansion. However, the CME center speed is almost identical to the speed of the solar wind, so no further deceleration due to drag is expected. Note that a deceleration of about 15\% in 24 hours is required to explain the CME speed profile by aging effects, which would be extreme for such a slow CME near 1~AU. 

The CME expansion with a speed of 30~km\,s$^{-1}$ corresponds to about $\Delta V_\mathrm{exp} / V_\mathrm{center}$ of 0.082. Using the slope of the velocity curve $\frac{\Delta V}{\Delta t} = 0.79$~m\,s$^{-2}$, the dimensionless expansion parameter of \citet{Demoulin:2008} can be calculated as $\xi  = 0.89$, which is slightly elevated, but not out of ordinary.

We were not able to identify a clear coronal counterpart to this magnetic ejecta from LASCO images. The shock corresponds to a relatively small increase in density followed right afterwards by a doubling of the density at a relatively sharp increase. The combination of the shock and post-shock compression resulted in a earthward displacement of Earth's magnetopause from 11.5 to 10~$R_E$. The magnetic ejecta resulted in an intense geomagnetic storm, whereas the sheath did not have any geo-effects in terms of storm or substorm.  
 
 %%%%%%%%%%%%%%%%%%%%%%%
\begin{figure*}[ht]
\centering
{\includegraphics*[width=8.5cm]{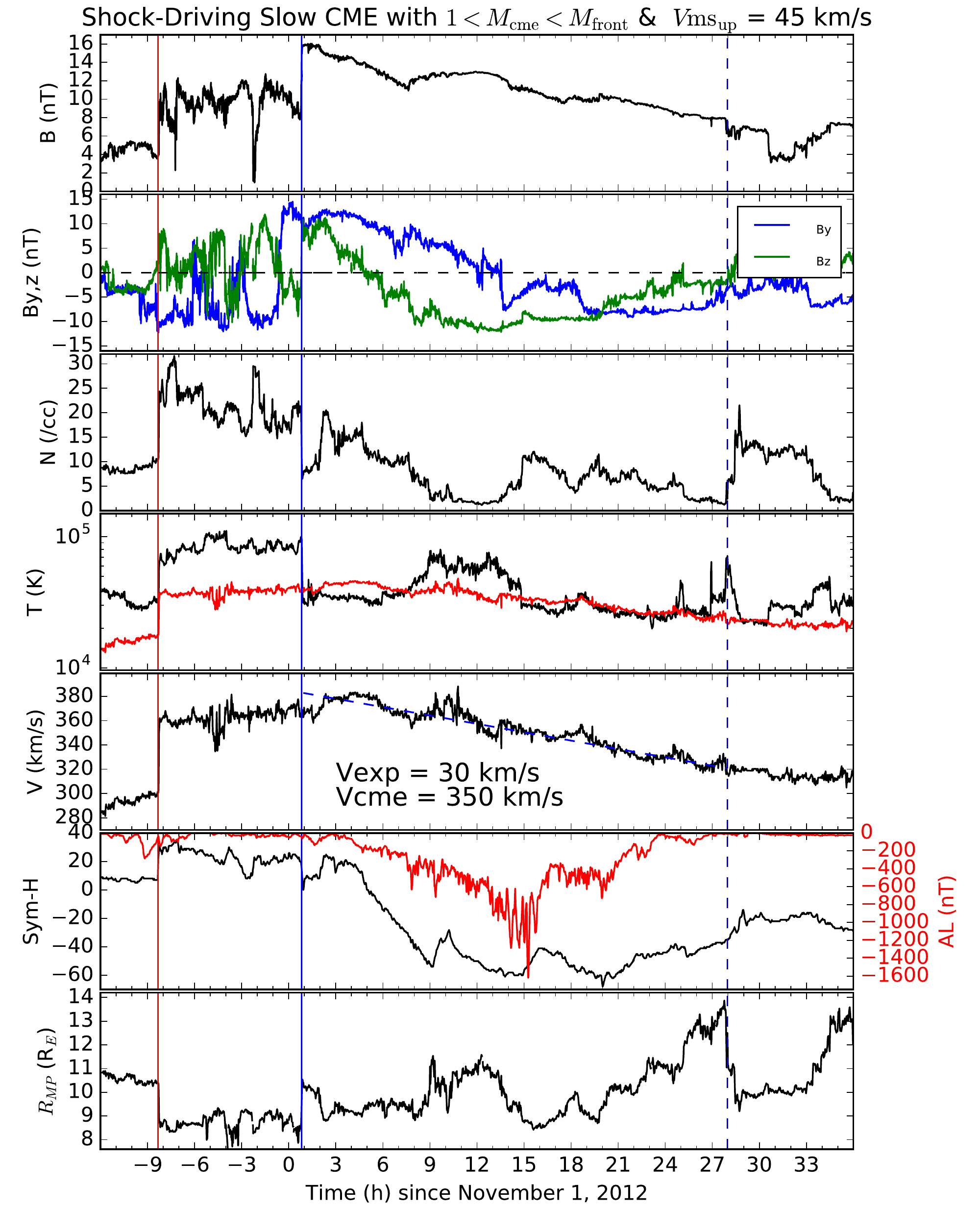}}
{\includegraphics*[width=8.5cm]{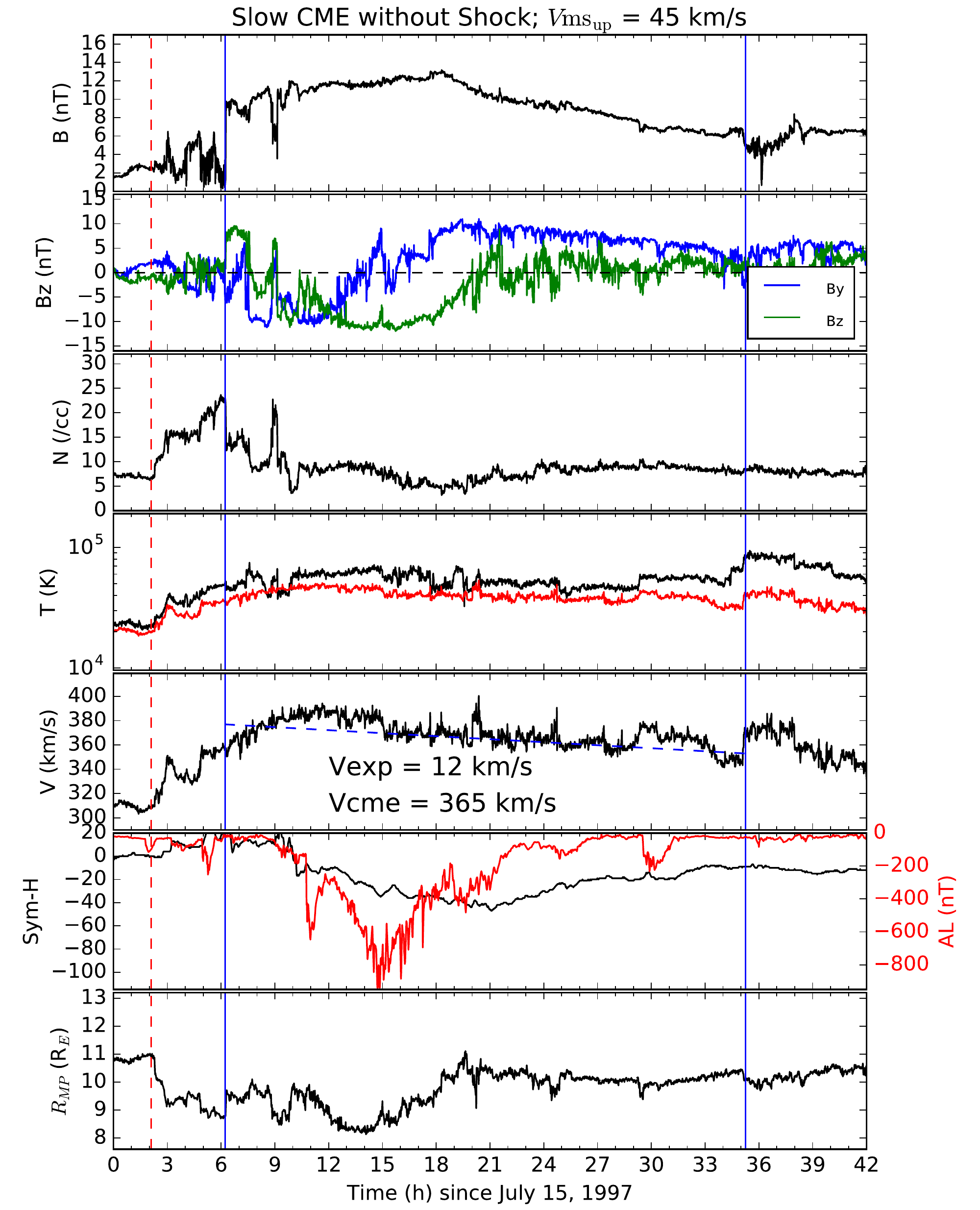}}
\caption{Same as Figure~\ref{Oct2001} but the left panels show the 2012 November 1 CME that drove a shock and the right panels the 1997 July 15 CME that did not drive a shock. The shock is marked with red vertical line, the magnetic ejecta boundaries with blue lines. The dashed red line indicates a compression feature marking the beginning of the swept-up sheath in front of the 1997 July magnetic ejecta.}
\label{Apr2001}
\end{figure*}
%%%%%%%%%%%%%%%%%%%%%%%%

\subsection{Example with $1 < M_\mathrm{cme} < M_\mathrm{front}$: 2012 October 31 (Event \#15)}

The 2012 November 1 magnetic ejecta (Figure~\ref{Apr2001}, left panels) is in many ways similar to the first example, where a slow solar wind speed and low fast magnetosonic speed made it possible for a slow but expanding CME to drive a fast-forward shock. Here, however, the CME center speed was fast enough to give rise to a shock by itself, although expansion played a role in enhancing the strength and increasing the speed of the shock. On 2012 October 31 at 16:55~UT, a fast-forward shock with speed of 390~km\,s$^{-1}$ and a Mach number of about 2.1 impacted Wind. Seven hours later around 00~UT, a decrease in temperature, density and an increase in magnetic field strength marked the beginning of a magnetic ejecta. The maximum speed of the ejecta, reached close to the front of the ejecta, was 380~km\,s$^{-1}$. The different CME databases give an end boundary between 02:30 UT and 05:00 UT on November 2, in close agreement. We use the boundary at 02:30 thereafter, but the different boundaries do not influence the results. The average CME speed is 350~km\,s$^{-1}$.  

The solar wind speed upstream of the shock was 290~km\,s$^{-1}$ and the fast magnetosonic speed was 45~km\,s$^{-1}$. The combination of very low solar wind speed and low fast magnetosonic speed resulted in this slow CME to have a center Mach number of 1.3, consistent with its driving a shock. The CME expansion resulted in a front speed of 380~km\,s$^{-1}$, which is consistent with the shock speed and strength. Although expansion is not essential to drive the shock, it made the shock faster and stronger. The expansion speed is 30~km\,s$^{-1}$ using the difference between the front velocity and the center speed of the CME; this gives a $\Delta V_\mathrm{exp} / V_\mathrm{center}$ of 0.085. Using the slope of the velocity curve $\frac{\Delta V}{\Delta t} = 0.64$~m\,s$^{-2}$, the dimensionless expansion parameter of \citet{Demoulin:2008} can be calculated as $\xi  = 0.78$, which is relatively typical. 

The shock compressed Earth's magnetopause from about 10.5 to 8.5~$R_E$ but the shock and the sheath did not result in significant geo-effects in terms of storm or substorm. The magnetic ejecta that drove the shock resulted in a moderate geomagnetic storm, as well as one relatively large substorm. 

In order to confirm that this CME's relatively low speed at 1~AU was not due to a strong heliospheric deceleration, we identified the associated eruption. There is only one full and one partial halo from October 26 12 UT to October 30 12 UT ({\it i.\ e}.\ 29 to 125 hours before the shock arrival time). The partial halo was a clear western limb event, which leaves the full halo on October 27 at 16:48 (96 hours before the shock arrival) as the best candidate for the CME measured {\it in situ}. STEREO/SECCHI images reveal that this was indeed a Earth-directed event. Second-order fit to the time-height plot in LASCO field-of-view gives an accelerating CME with a leading edge speed around 24~$R_\odot$ of 500~km\,s$^{-1}$. \citet{Hess:2017} identified this eruption as the source of the CME and shock measured {\it in situ} and listed a fitting 3-D speed of 373~km\,s$^{-1}$ for the CME based on coronagraphic observations and the Graduated Cylindrical Shell method of \citet{Thernisien:2009}. HELCATS reports heliospheric observations of this CME by both STEREO-A and STEREO-B, with STEREO-A providing observations up to elongations of 40$^\circ$. Fitting to the time-elongation plots give a heliospheric speed between 396 and 471~km\,s$^{-1}$ for STEREO-A, depending on the method chosen. Overall, this confirms that this CME was relatively slow in the corona and in the heliosphere with a likely speed between 400 and 500~km\,s$^{-1}$. {\it In situ} maximum speeds around 380~km\,s$^{-1}$ and the arrival time are consistent with a CME speed in the corona of about 500~km\,s$^{-1}$ and an heliospheric speed around 450~km\,s$^{-1}$.

\subsection{Counter-Example: Slow Non-Shock Driving CME: 1997 July 15}

The right panel of Figure~\ref{Apr2001} shows the  {\it in situ} measurements associated with a slow CME that impacted Wind on 1997 July 15. 
This event is in many ways similar to the first and third discussed events in terms of magnetic field strength ($B_\mathrm{max} \sim 13$~nT), slow upstream solar wind speed below 320~km\,s$^{-1}$, low upstream fast magnetosonic sped of about 45~km\,s$^{-1}$, and CME speed of about 365~km\,s$^{-1}$.  In fact, under such conditions, that CME would be expected to drive a weak shock, as the CME center Mach number is 1.1. However, the CME is not expanding, and in fact the CME front, with a speed under 360~km\,s$^{-1}$ is slower than the center. As such, it does not drive a shock. However, there is a clear swept-up sheath in front of the magnetic ejecta starting about 4 hours before the start of the ejecta and characterized by higher proton densities and a more turbulent magnetic field. This sheath starts with a jump in density, velocity in temperature, similar but not as steep as a shock. The CME expansion is about 12~km\,s$^{-1}$, giving a  $\Delta V_\mathrm{exp} / V_\mathrm{center} \sim 0.03$, {\it i.\ e.} low. Using  the velocity slope value of $\frac{\Delta V}{\Delta t} = 0.23$~m\,s$^{-2}$, the dimensionless expansion parameter is found to be $\xi  = 0.26$, indicating a very weak expansion. Note that this low expansion does not appear to be related to interaction with another CME or a fast solar wind stream, as the solar wind behind the CME had a speed between 360 and 380~km\,s$^{-1}$. The Dst index reached $-45$~nT during the magnetic ejecta, and the dense, swept-up sheath compressed the magnetopause from 11 to 9~$R_E$. \citet{Lugaz:2016a} analyzed in details {\it in situ} measurements associated with another slow non-shock driving CME that was preceded by a longer and denser swept-up sheath on 2013 January 17. 

We do not give an example of a complex CME that nonetheless drove a shock. The 2012 September 30 CME was studied in details in \citet{Liu:2014b} and \citet{Liu:2016} and interested readers can refer to this example of an overtaken slow CME driving a shock.

%%%%%%%%%%%%%%%%%%%%%%%%%%%%%%%%%%%%%%%%%%%%%%%%%%%%%%%%%%%%%%%%%%%%%%%%%%%%
\section{Proportion of CMEs Driving Shocks: Dependence on CME Speed} \label{data}
%%%%%%%%%%%%%%%%%%%%%%%%%%%%%%%%%%%%%%%%%%%%%%%%%%%%%%%%%%%%%%%%%%%%%%%%%%%%%
Obviously, the proportion of CMEs that drive shocks should increase with increasing CME speeds. In addition, CME expansion speed is positively correlated with CME central speed \citep[]{Richardson:2010}, while correlation analysis shows that the expansion speed is proportional to the square of the CME central speed \citep[]{Gulisano:2010}. Therefore, fast CMEs have, on average, large expansion speeds and fast leading edges. This general expectation is, of course, made more complex by the time-varying and non-uniform nature of the solar wind and interplanetary magnetic field. As such, a 500~km\,s$^{-1}$ CME behind a fast solar wind stream may not drive a shock, whereas, as shown above, some CMEs with speeds of 360~km\,s$^{-1}$ drive shocks if the conditions are right (low solar wind speed, among others). It has been reported that solar cycle 24 (SC24) has a lower average solar wind speed and lower interplanetary magnetic field as compared to SC23. Thus, we might expect that slower CMEs may be more likely to drive a shock in SC24 than in SC23. We attempt to test this hypothesis in this section.

\begin{table*}[t]
\centering
\begin{tabular}{|c|c|ccc|cc|}
\hline
Period & $V_\mathrm{sw}$ & $V_a$  & $C_{s1}$ & $C_{s2}$ & $V_\mathrm{min1}$ & $V_\mathrm{min2}$\\
\hline
\multicolumn{7}{|c|}{{\bf Median}}\\
\hline
05/1996 - 2007 & 420 & 56.3 & 55.3 & 47.6 & 502 & 497 \\ % done
2008 - 01/2017 & 394 & 48.6 & 53.1 & 42.3 & 470 & 463 \\ % done
\hline
07/1997 - 01/2006 & 426 & 61.7 & 55.8 & 48.9 & 512 & 507 \\% done
10/2010 - 11/2016 & 399 & 51.7 & 53.5 & 43.3 & 477 & 471 \\% done
\hline
\multicolumn{7}{|c|}{{\bf Average}}\\
\hline
05/1996 - 2007 & 442 & 60.3 & 57.2 & 50.0 & 527 & 522 \\ % done
2008 - 01/2017 & 415 & 50.9 & 55.1 & 45.1 & 491 & 484 \\ % done
\hline
07/1997 - 01/2006 & 448 & 66.1 & 57.9 & 51.6 & 538 & 534 \\% done
10/2010 - 11/2016 & 416 & 54.7 & 55.3 & 45.9 & 495 & 489 \\% done
\hline
\end{tabular}
\label{coucouc}
\caption{Different velocities obtained from 1-day OMNI data: solar wind, Aflv{\'e}n, ion-acoustic (two values for the electron temperature) and minimum speed to drive a shock. All velocities are in km\,s$^{-1}$. $V_\mathrm{min}$ is the sum of the solar wind and fast magnetosonic speeds.}
\end{table*}

\subsection{General Considerations and Expectations}
We use 1-day OMNI data to obtain the solar wind speed and the characteristic speed for the past 21 years. OMNI does list the solar wind Mach number but not the fast magnetosonic speed. It can however be derived from the proton density, temperature and the magnetic field measured near 1~AU. 
We neglect the influence of alpha particles, which result in a slightly lower Alfv{\'e}n speed, $V_A$, compared to the one used in OMNI to derive the Mach number. Because electron temperature is not part of OMNI, we use two assumptions to calculate the ion-acoustic speed, $V_s$: $T_p = T_e$ and $T_e = 1.4 \times 10^5$~K. The second assumption is the one used in OMNI to derive their Mach number. We also neglect the influence of alpha particles in the sound speed. For the fast magnetosonic speed, we use the formula $V_\mathrm{ms} = \sqrt{V_A^2 + V_s^2}$. 

Our primary quantity of interest is the sum of the solar wind speed and the fast magnetosonic speed. This represents the minimum front speed a transient needs to have to be able to drive a shock, and we refer to it as $V_\mathrm{min}$. We also compare the values of the ion-acoustic speed, the Alfv{\'e}n speed and the solar wind speed to determine where the difference between solar cycles originate from. We use May 1996 -- December 2007 for SC23 and January 2008 -- January 2017 for SC24. 

\subsection{Comparing Solar Cycles 23 and 24}
The average $V_\mathrm{min}$ for SC23 and SC24 are 527~km\,s$^{-1}$ and 491~km\,s$^{-1}$, respectively. This means that a 500~km\,s$^{-1}$ CME can be expected, on average, to drive a shock in SC24 but not in SC23. Table~3 %\ref{coucouc} 
lists the different values of the speeds for the two solar cycles. For SC24, we use the time period from January 2008 to January 2017, inclusive. It can be seen that the main differences are due to the background solar wind speed, accounting for about 27~km\,s$^{-1}$ of difference, and the Alfv{\'e}n speed, accounting for about 9~km\,s$^{-1}$ of difference. The difference is approximately independent of the choice of $T_e$ for the ion-acoustic speed.

We looked more closely only at periods when CMEs occurred, as the difference found between SC23 and SC24 may arise from the prolonged solar minimum between the two cycles, a time when there were no CMEs, and for which a low solar wind speed would have no influence on the formation of shocks in front of slow CMEs. We limit ourselves to time periods when the 13-month smoothed sunspot number was above 30. This corresponds to the time period from July 1997 to January 2006 for SC23 and October 2010 to November 2016 for SC24. Considering only these two time periods also allows us to control for the fact that SC24 is still ongoing. The average $V_\mathrm{min}$ with these restrictions is 538~km\,s$^{-1}$ for SC23 as compared to 495~km\,s$^{-1}$ for SC24. The difference between SC23 and SC24 is statistically significant, both for the full cycles as well as the active period only. It is unclear whether CME periods should be removed from these statistics. On the one hand, CMEs have typically higher speeds and higher fast magnetosonic speeds than the background solar wind and SC23 had many more fast CMEs with strong magnetic field measured at L1 as compared to SC24. On the other hand, CMEs constitute part of the medium into which following CMEs propagate and shocks often form inside previous CMEs \citep[]{Collier:2007, Lugaz:2015a, Lugaz:2016b}. Removing days for which the magnetic field strength was above 10~nT has almost no influence on the values of $V_\mathrm{min}$ and therefore we do consider this question more closely. 

\begin{figure}[b]
\centering
{\includegraphics*[width=7.5cm]{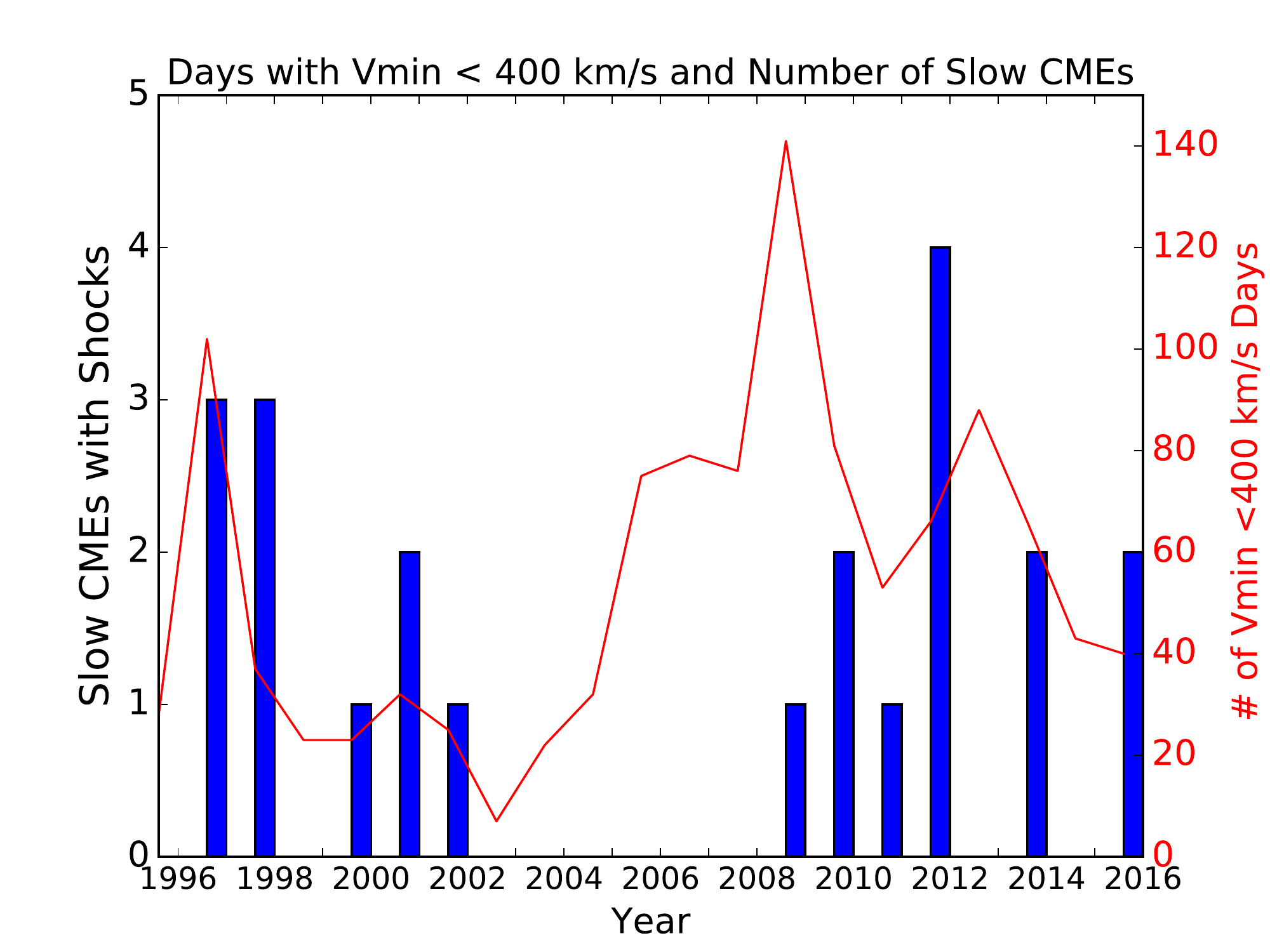}}
\caption{Yearly distribution of the number of slow CMEs (with average speed less or equal to 370~km\,s$^{-1}$) with shocks and of the number of days with $V_\mathrm{min} <$ 400~km\,s$^{-1}$}
\label{noshock}
\end{figure}

For periods with the sunspot number greater than 30, we also looked at the number of days when the minimum speed was low. SC23 had 262 days with $V_\mathrm{min} <$ 400~km\,s$^{-1}$ and a median of 512~km\,s$^{-1}$. SC24 had 382 days with $V_\mathrm{min} <$ 400~km\,s$^{-1}$ and a median of 477~km\,s$^{-1}$. Note this difference occurs even though SC23 had 878 more days with sunspot number greater than 30 than SC24. Further results are shown in Table~4 %\ref{days}
 and Figure~\ref{noshock}.

\begin{table*}[t]
\centering
\begin{tabular}{|c|c|c|c|c|}
\hline
Period & \# Days & $V_\mathrm{min}<$350 & $V_\mathrm{min}<$400 & $V_\mathrm{min}<$450 \\
\hline
 07/1997 - 01/2006 & 3128 & 19 & 262 & 764 \\
 10/2010 - 11/2016 & 2250 & 60 & 382 & 838\\
 \hline
\end{tabular}
\label{days}
\caption{Number of days with low $V_\mathrm{min}$ during periods with greater than 30 monthly sunspot number. We exclude a dozen days with data gap.}
\end{table*}

%19 days with $V_\mathrm{min} <$ 350~km\,s$^{-1}$, 263 days with $V_\mathrm{min} <$ 400~km\,s$^{-1}$, and  765 days with $V_\mathrm{min} <$ 450~km\,s$^{-1}$, for a median of 512~km\,s$^{-1}$. SC24 had 60 days with $V_\mathrm{min} <$ 350~km\,s$^{-1}$, 383 days with $V_\mathrm{min} <$ 400~km\,s$^{-1}$, and 840 days with $V_\mathrm{min} <$ 450~km\,s$^{-1}$, for a median of 477~km\,s$^{-1}$. Note this difference occurs even though SC23 had 875 more days with sunspot number greater than 30 than SC24. 

Based on these results, our expectations are that CMEs with front speeds at 1~AU under 500~km\,s$^{-1}$ are more likely to drive a shock in SC24 than in SC23. We test this expectation in the next subsection.

  %%%%%%%%%%%%%%%%%%%%%%%
%%%%%%%%%%%%%%%%%%%%%%%%

\begin{table*}[ht]
\centering
\begin{tabular}{|c|c|c|c|c|c|}
\hline
Speed & Total \# of CMEs & \multicolumn{2}{|c|}{ 05/1996 - 2007} & \multicolumn{2}{|c|}{2008- 11/2016}\\
\hline
& & CME with shock & \% & CME with shock & \% \\
 \hline
$<$ 370 &  74 & 9/33 & 27\% & 11/41 & 27\%\\
370-390 & 73 & 12/35 & 34\% & 12/38 & 32\%\\
400-420 & 71 & 21/48 & 44\% & 12/23 & 52\%\\
430-450 & 83 & 24/57 & 42\% & 11/26 & 42\%\\
460-500 & 72 & 24/46 & 52\% & 14/26 & 54\%\\
\hline
\end{tabular}
\label{ShockStats}
\caption{Number and percentage of CMEs with shocks for a given range of CME speeds for solar cycle 23 and 24.}
\vspace{0.4cm}
\end{table*}

\subsection{Proportion of CMEs with Shocks}
We use the ICME database of \citet{Richardson:2010} combined with the shock database of \citet{Kilpua:2015}. We associate shocks with ICMEs based on the shock time being within 2 hours of the ICME disturbance start time as listed by \citet{Richardson:2010}. We focus only on CMEs with average speed under 500~km\,s$^{-1}$, based on the statistical considerations presented in the previous subsection. There were 373 such CMEs over these two solar cycles: 219 in SC23 and 154 in SC24. Hereafter, we refer to them as slow CMEs. Comparing the proportion of slow CMEs with shocks over these two solar cycles is not straightforward, since the median speed of a slow CME in solar cycle 24 is 390~km\,s$^{-1}$ but it is 410~km\,s$^{-1}$ in solar cycle 23. We compared the proportion of CMEs with or without shocks in five speed ranges, as well as the cumulative proportion for different speeds.

\begin{figure}[b]
\centering
{\includegraphics*[width=7.5cm]{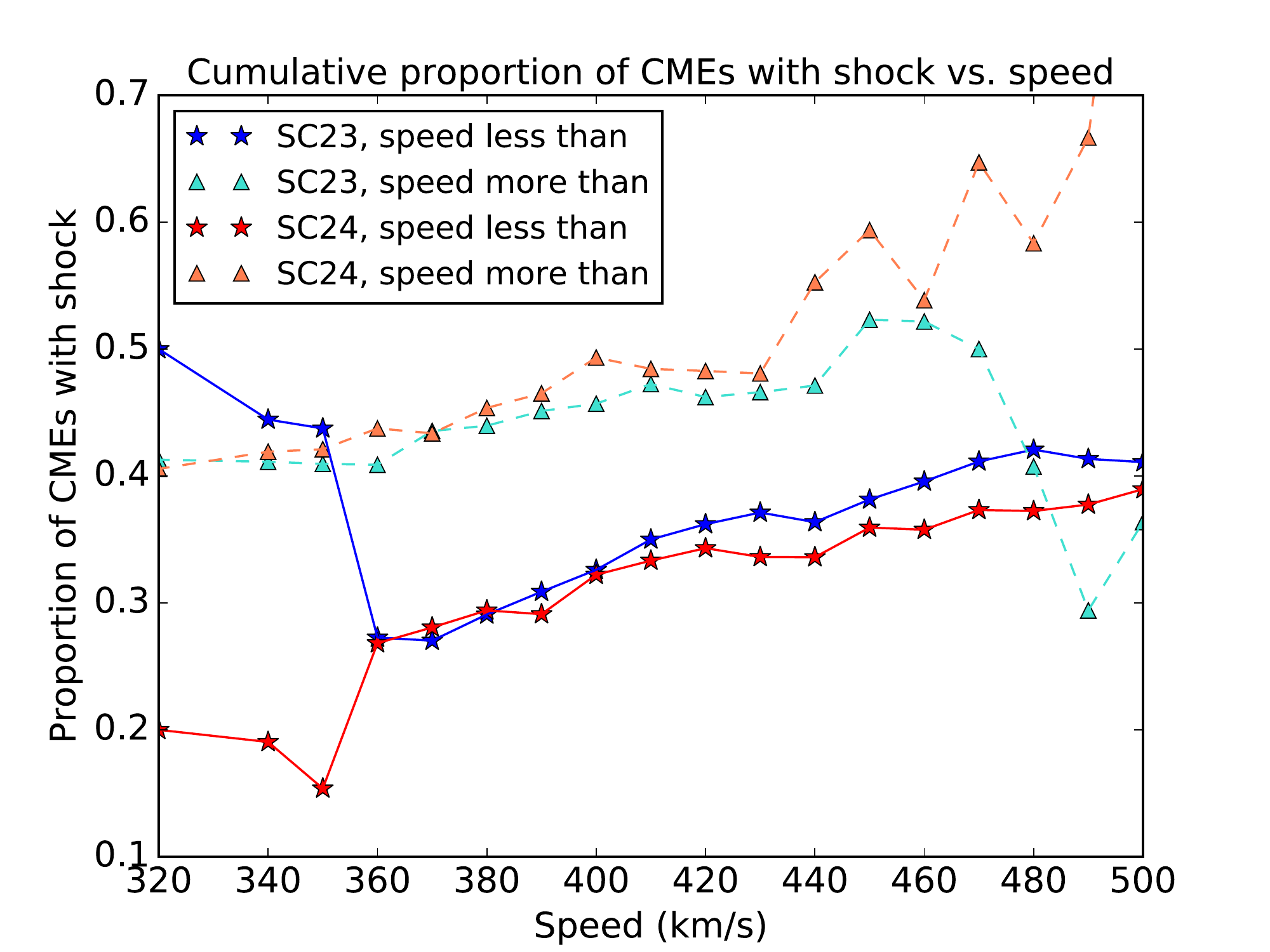}}
\caption{Cumulative proportion of CMEs with shocks for increasing speeds (solid) and decreasing speed (dashed). Blue is for solar cycle 23 and red for solar cycle 24. This shows that, for example, using the solid curves that 29\% of CMEs with speed below 380~km\,s$^{-1}$ drive a shock and, using the the dashed curves, that 48\% of CMEs with speed between 410 and 500 km\,s$^{-1}$ drive a shock.}
\label{cumulative}
\end{figure}

Overall, 27\% of CMEs with average speed less or equal to 370~km\,s$^{-1}$ drive a shock. The proportion does not change from SC3 to SC24. 41\% of CMEs with an average speed below 500~km\,s$^{-1}$ drive a shock in solar cycle 23 {\it vs.} 39\% in solar cycle 24. Table~5 %\ref{ShockStats}
 shows the statistics and Figure~\ref{cumulative} show the cumulative proportion of CMEs with shocks given the CME speed. It is clear that there is almost no difference between the two solar cycles. 

The table illustrates that about one quarter of CMEs with speed below 370~km\,s$^{-1}$, and about half of CMEs with speeds between 460 and 500~km\,s$^{-1}$ drive a shock. In Figure~\ref{cumulative}, the solid curves should be read from left to right (for example, 29\% of CMEs with speed below 380~km\,s$^{-1}$ drive a shock) and the dashed curves from right to left (48\% of CMEs with speed between 410 and 500~km\,s$^{-1}$ drive a shock). This shows that as the CME speed increases, the percentage of CMEs with shocks increases, as expected. Except for the low statistics at the end of each curve, the results for SC23 and SC24 are almost identical. If anything, SC23 had an anomalously high percentage of slow CMEs with shocks (7 out 16 of CMEs with speed of 350~km\,s$^{-1}$ or less drove a shock)  and an anomalously low percentage of fast CMEs with shocks (5 of 17 CMEs with speeds of 490 or 500~km\,s$^{-1}$ drove a shock). We do not consider speeds higher than 500~km\,s$^{-1}$ because CME-CME interaction and data gaps would need to be taken more carefully into account.  Note that, in this section, we use the average CME speed as listed in \citet{Richardson:2010} whereas in the previous section, we recalculated it for all CMEs with listed speed less than 380~km\,s$^{-1}$. We also excluded some CMEs based on the abnormal sheath duration or the presence of multiple shocks. For these reasons, we find 20 CMEs with speed less than 370~km\,s$^{-1}$ that drove a shock in this section but 21 (and one with a speed of 375~km\,s$^{-1}$) in the previous section.

\section{Discussions and Conclusion} \label{conclusion}

SC24 has a statistically significant lower threshold of CME speed for a CME to drive a shock as compared to SC23. The difference of about 35~km\,s$^{-1}$ corresponds to about half the fast magnetosonic speed at 1~AU, so it is non-negligible when considering the ability of a CME to drive a shock. The fact that the proportion of slow CMEs driving shocks in SC4 is almost unchanged from that of SC23 for a given speed range means that there must be another  factor compensating for the lower $V_\mathrm{min}$ in SC24. We hypothesize that this may be CME expansion. 

\citet{Gopalswamy:2014}, (\citeyear{Gopalswamy:2015}) reported on anomalous CME expansion in SC24. \citet{Gopalswamy:2015}, who focused on magnetic clouds (MCs), found that MCs in SC24 are slower on average than those in SC23. Interestingly, the expansion speed was found to be smaller by about a factor of 2 (from 51~km\,s$^{-1}$ to 25~km\,s$^{-1}$). A similar value for SC23 was reported in \citet{Lepping:2008b}. The MC sizes were also smaller in SC24 than in SC23. The reduction in the MC speed as well as the expansion speed result in the dimensionless expansion parameter, following \citet{Demoulin:2009}, to be more or less constant between the two solar cycles. However, it is clear that this difference in expansion velocity means that the leading edge of a CME of a given average speed is lower in SC24 than it was in SC23. Note that a more in-depth analysis of expansion speed is required: (i) \citet{Gopalswamy:2015} only considered the front-to-back difference in speed to determine the expansion speed, whereas \citet{Demoulin:2009} uses the velocity slope through the unperturbed part of the ejecta to calculate the dimensionless expansion parameter, (ii) the study of \citet{Gopalswamy:2015} needs to be expanded to non-MC magnetic ejecta, which will greatly increase the statistics (from about 100 events in that study to 400 events if studying all CMEs in SC23 and SC24). 

As discussed in \citet{Gopalswamy:2015}, the changes in size and expansion speed but not in dimensionless expansion parameter can be explained by a wider CME size close to the Sun \citep[]{Gopalswamy:2014,Gopalswamy:2015c} and an ``over-expansion'' up to larger distances but a similar evolution of the CME size with distance \citep[]{Gopalswamy:2015} in the heliosphere. Work using Messenger data \citep[]{Winslow:2015} has shown that the decrease of the magnetic field inside magnetic ejecta with distance between 0.4 and 1~AU is approximately the same in solar cycle 24 as it was in solar cycle 21 with Helios data \citep[]{Bothmer:1998, Liu:2005}. Further work is required to confirm the existence and cause of a lower CME expansion speed at 1~AU in SC24.

  %%%%%%%%%%%%%%%%%%%%%%%
%%%%%%%%%%%%%%%%%%%%%%%%

We found that, even for the slowest CME speeds, about one quarter of CMEs drive shocks. It can appear surprising that this number does not go towards 0 even for CMEs with speed below 340~km\,s$^{-1}$. There could be at least three different explanations: (i) there is a bias towards identifying shock-driving CMEs, (ii) many non-shock driving, slow CMEs fully reconnect before impacting Earth, or (iii) expansion contributes to slow CMEs driving shocks. A bias towards identifying  shock-driving slow CMEs {\it vs.} non-shock-driving slow CMEs is possible, but unlikely to explain this entire difference. The database contains 30 CMEs with average speed less or equal to 340~km\,s$^{-1}$, 8 of which drove a shock, and there would need to be an additional 20 or more unidentified non-shock driving CMEs for the proportion to be significantly different. CME erosion has been reported before \citep[]{Ruffenach:2012}, and it is possible that the presence of a shock and sheath region where the flows are deflected ``protects'' the magnetic ejecta from being eroded. However, there has not been previously reported studies about this. The importance of expansion for CMEs to drive shocks has been reported, as discussed in the introduction, for observations at high latitudes \citep[]{Gosling:1994}, discussed for CMEs near 1~AU by \citet{Siscoe:2008}, and reported in a numerical simulation by \citet{Poedts:2016}. 

Based on the current study and these previous works, we propose the following scenario for these shock-driving slow CMEs. A relatively slow CME erupts from the Sun with a speed between 350 and 500~km\,s$^{-1}$, or about 50~km\,s$^{-1}$ faster than the solar wind speed at 1~AU. Such slow CMEs do not drive shocks in the corona, where the fast magnetosonic speed is greater than 200~km\,s$^{-1}$. Due to the small gradient in speed between the CME and the solar wind, the CME does not decelerate much as it propagates. Such a CME may reach Earth with a speed 20-30~km\,s$^{-1}$ faster than the solar wind speed. This would not be enough to drive a shock. However, CME expansion adds another 40-50~km\,s$^{-1}$ to the speed of the leading edge of the CME, making the CME front super-fast and causing a shock. For these CMEs, expansion contributes between 40 and 80\% to the formation of a shock. By this, we mean that the CME front speed in the solar wind frame is caused for 40 to 80\% by expansion, the rest being due to the CME propagation. This number is larger than the 35\% reported by \citet{Owens:2005} for fast CMEs. Some of these shock-driving slow CMEs are convected with the solar wind but they are ``overtaking'' a slower solar wind stream. Such CMEs may be caused by streamer blowouts near the Sun. In SC4, the reduced expansion speed compensates the reduced solar wind and magnetosonic speeds, resulting in a more or less constant proportion of CMEs with shocks in SC24 as compared to SC23.

\subsection{Conclusions}

We identified the slowest CMEs that drove a shock from 1996 to 2016, focusing on 22 events with an average speed under 375~km\,s$^{-1}$. For eleven of these events, the CME average speed is sub-fast in the solar wind frame, but the CME front speed, taking into consideration the CME radial expansion, is super-fast or with a Mach number close to unity. Although the CME radial expansion always occurs sub-Alfv{\'e}nically, we find that CME expansion is central in explaining why these slow CMEs drove a shock. Other conditions must be met for these CMEs to drive shocks, including slow upstream solar wind speed (typically around 310~km\,s$^{-1}$) and relatively slow upstream fast magnetosonic speed (around 55~km\,s$^{-1}$). These conditions correspond to the 10\% lowest solar wind speeds in the past 21 years. For such slow CMEs, all conditions must be met (significant expansion, low solar wind speed and low upstream magnetosonic speed). This investigation should be extended to CMEs of all speeds, with or without a shock,  to determine the importance of expansion in the driving and strengthening of shocks. The current study has revealed that such an in-depth investigation is warranted, since for some CMEs, radial expansion has now been found to be important for the CME ability to drive a shock.

We also determined the percentage of CMEs that drive shocks depending on the CME average speed. We find that, even for the lowest CME speeds, the proportion of CMEs that drive shocks remains around 25\% and that proportion increases to 50\% for CMEs with speeds between 460 and 500~km\,s$^{-1}$. The non-negligible percentage of slow CMEs that drive shocks can be best understood as originating from the influence of CME expansion in having the CME leading edge moving super-Alfv{\'e}nically in the solar wind frame. 

Lastly, in this initial study, we found evidence that the percentage of CMEs that drive shocks, once the CME speed is taken into consideration, does not change between SC23 and SC24. This occurs, even though we find that the average minimum required speed for a CME to drive a shock in SC24 was about 30--35~km\,s$^{-1}$ lower than it was in SC23. Such a difference should have had an influence in the percentage of CMEs with shocks, as it is comparable to half the typical fast magnetosonic speed at 1~AU. One likely explanation of this lack of increase percentage of CMEs with shocks is the decrease in CME expansion speed that was reported at 1~AU for magnetic clouds by \citet{Gopalswamy:2015}. Further investigations of CME expansion and its difference between SC23 and SC24 are required to further validate this hypothesis.

\begin{acknowledgments}

The research for this manuscript was supported by the following grants: NSF AGS-1435785 and AGS-1433213 and NASA NNX15AB87G and NNX16AO04G. N.~L. was also partially supported by NASA RBSP ECT contract to UNH. N.~L. and C.~J.~F were also partially supported by NASA grant NNX15AU01G.
\end{acknowledgments}

\bibliographystyle{apj}

\end{document}